\documentclass[11pt]{article}
\usepackage{times}
\usepackage{graphicx}

\makeatletter
\def\maketitle{\par 
\begingroup
   \def\thefootnote{\fnsymbol{footnote}}
   \def\@makefnmark{\hbox to 0pt{$^{\@thefnmark}$\hss}} % for perfect author name centering
   \long\def\@makefntext##1{\parindent 1em\noindent \hbox to1.8em{\hss $\m@th ^{\@thefnmark}$}##1}
   \@maketitle \@thanks
\endgroup
\setcounter{footnote}{0}
\let\maketitle\relax \let\@maketitle\relax
\gdef\@thanks{}\gdef\@author{}\gdef\@title{}\let\thanks\relax}
\makeatother

\makeatletter
\def\@maketitle{\vbox{\hsize\textwidth
\linewidth\hsize \vskip 0.1in \toptitlebar \centering
{\LARGE\bf \@title\par}  \bottomtitlebar % \vskip 0.1in %  minus
   \def\And{\end{tabular}\hfil\linebreak[0]\hfil
            \begin{tabular}[t]{c}\bf\rule{\z@}{24pt}\ignorespaces}% 
   \def\AND{\end{tabular}\hfil\linebreak[4]\hfil
            \begin{tabular}[t]{c}\bf\rule{\z@}{24pt}\ignorespaces}% 
   \def\LINEBREAK{\end{tabular}\linebreak[4]\begin{tabular}[t]{c}\bf\rule{\z@}{16pt}\ignorespaces}% 
    \begin{tabular}[t]{c}\bf\rule{\z@}{24pt}\@author\end{tabular}% 
\vskip 0.3in minus 0.1in}}
\makeatother

\renewenvironment{abstract}{\vskip.075in\centerline{\large\bf Abstract}\vspace{0.5ex}\begin{quote}}{\par\end{quote}\vskip 1ex}

\makeatletter
\def\section{\@startsection {section}{1}{\z@}{-2.0ex plus -0.5ex minus -.2ex}{1.5ex plus 0.3ex minus0.2ex}{\large\bf\raggedright}}
\def\subsection{\@startsection{subsection}{2}{\z@}{-1.8ex plus-0.5ex minus -.2ex}{0.8ex plus .2ex}{\normalsize\bf\raggedright}}
\def\subsubsection{\@startsection{subsubsection}{3}{\z@}{-1.5ex plus -0.5ex minus -.2ex}{0.5ex plus .2ex}{\normalsize\bf\raggedright}}
\def\paragraph{\@startsection{paragraph}{4}{\z@}{1.5ex plus 0.5ex minus .2ex}{-1em}{\normalsize\bf}}
\def\subparagraph{\@startsection{subparagraph}{5}{\z@}{1.5ex plus  0.5ex minus .2ex}{-1em}{\normalsize\bf}}

\makeatother

\skip\footins 9pt plus 4pt minus 2pt
\def\footnoterule{\kern-3pt \hrule width 12pc \kern 2.6pt }
\leftmargini\leftmargin \leftmarginii 2em
\makeatletter
\def\@listi{\leftmargin\leftmargini}
\def\@listii{\leftmargin\leftmarginii
   \labelwidth\leftmarginii\advance\labelwidth-\labelsep
   \topsep 2pt plus 1pt minus 0.5pt
   \parsep 1pt plus 0.5pt minus 0.5pt
   \itemsep \parsep}
\def\@listiii{\leftmargin\leftmarginiii
    \labelwidth\leftmarginiii\advance\labelwidth-\labelsep
    \topsep 1pt plus 0.5pt minus 0.5pt 
    \parsep \z@ \partopsep 0.5pt plus 0pt minus 0.5pt
    \itemsep \topsep}
\def\@listiv{\leftmargin\leftmarginiv
     \labelwidth\leftmarginiv\advance\labelwidth-\labelsep}
\def\@listv{\leftmargin\leftmarginv
     \labelwidth\leftmarginv\advance\labelwidth-\labelsep}
\def\@listvi{\leftmargin\leftmarginvi
     \labelwidth\leftmarginvi\advance\labelwidth-\labelsep}
\makeatother

\belowdisplayskip \abovedisplayskip
\makeatletter
\def\normalsize{\@setsize\normalsize{11pt}\xpt\@xpt}
\def\small{\@setsize\small{10pt}\ixpt\@ixpt}
\def\footnotesize{\@setsize\footnotesize{10pt}\ixpt\@ixpt}
\def\scriptsize{\@setsize\scriptsize{8pt}\viipt\@viipt}
\def\tiny{\@setsize\tiny{7pt}\vipt\@vipt}
\def\large{\@setsize\large{14pt}\xiipt\@xiipt}
\def\Large{\@setsize\Large{16pt}\xivpt\@xivpt}
\def\LARGE{\@setsize\LARGE{20pt}\xviipt\@xviipt}
\def\huge{\@setsize\huge{23pt}\xxpt\@xxpt}
\def\Huge{\@setsize\Huge{28pt}\xxvpt\@xxvpt}
\makeatother

\def\toptitlebar{\hrule height4pt\vskip .25in\vskip-\parskip}
\def\bottomtitlebar{\vskip .29in\vskip-\parskip\hrule height1pt\vskip .09in}
\begin{document}

\title{Information transmission in oscillatory\\ neural activity}
\author{Kilian Koepsell \& Friedrich T. Sommer\\
Redwood Center for Theoretical Neuroscience \\
Helen Wills Neuroscience Institute \\
University of California, Berkeley \\
Berkeley, CA 94720 \\
\texttt{\textrm\{kilian,fsommer\textrm\}@berkeley.edu}}
\maketitle

\begin{abstract}
  Periodic neural activity not locked to the stimulus or to motor responses is
  usually ignored. Here, we present new tools for modeling and quantifying the
  information transmission based on periodic neural activity that occurs with
  quasi-random phase relative to the stimulus. We propose a model to reproduce
  characteristic features of oscillatory spike trains, such as histograms of
  inter-spike intervals and phase locking of spikes to an oscillatory
  influence. The proposed model is based on an inhomogeneous Gamma process
  governed by a density function that is a product of the usual
  stimulus-dependent rate and a quasi-periodic function.  Further, we present
  an analysis method generalizing the direct method
  \cite{Rieke1999,Brenner2000} to assess the information content in such
  data. We demonstrate these tools on recordings from relay cells in the
  lateral geniculate nucleus of the cat.
\end{abstract}
%\begin{keyword}axonal delays, synchronization, zero-phase-lag, efficacy suppression
%\end{keyword}
%\end{frontmatter}

\section{Introduction}
\label{intro}

Oscillatory activity is ubiquitous in the brain, manifesting itself on a
macroscopic level in EEG and MEG recordings as ``brain waves'' and on the
level of single neurons in micro-electrode recordings as periodic spike
patterns. There are different forms of oscillations. Traditionally one
distinguishes oscillations that co-occur with a behavioral condition such as
sensory stimulation from ongoing oscillations that are spontaneous, that is,
present independently of the behavioral condition. When co-occurring with
stimulation, two different types of oscillations have been discerned: {\it
  Stimulus-evoked} oscillations are phase-locked to the stimulus, that is, the
phase of the periodicity in neural activity can be reproduced by repeating the
same stimulus. In contrast, in {\it stimulus-induced} oscillations the phase
is variable with respect to the stimulus and cannot be predicted from one
stimulus trial to the next.

In various sensory systems, it has been shown that neural oscillations that
co-occur with stimulation play an important role for the coding of sensory
information. Examples are the olfactory system in species ranging from insects
to mammals~\cite{Adrian1942,Freeman1972,Gelperin1990,Laurent1994}, the whisker
system in rats~\cite{Szwed2003} and the somatosensory system in
primates~\cite{Ahissar1990}. The motivation for our present study is the
question whether oscillations that are spontaneous or not phase-locked to the
stimulus can have roles in information coding and transmission. This is
conceivable since ongoing or non-stimulus-locked oscillations could still be
influenced by the stimulus and thereby carry stimulus information. However, to
date, this question has not been satisfactory addressed. For example, consider
the visual system. Over the last four decades, numerous studies of various
vertebrate species have reported oscillatory activity in the early visual
system~\cite{Heiss1966,Laufer1967,Rodieck1967,Neuenschwander1996,Castelo-Branco1998,Ishikane2005,Montemurro2008}.
These oscillations are sometimes stimulus-evoked but often spontaneous. The
question of whether they are systematically influenced by the visual stimulus
and what their functional role could be, has been subject of controversy and
is still not resolved. One difficulty with addressing this question is the
lack of applicable models and analysis tools. To fill this gap, we propose two
new theoretical tools for studying the function of oscillatory activity: A
spiking neuron model that reproduces the characteristic statistical properties
of oscillatory neurons and a method to quantify the information of oscillatory
spike trains even in the absence of stimulus-locking.

One possibility how ongoing oscillations can be used to convey stimulus
information is in phase coding schemes, in which relative phases encode
information. A prominent example is phase coding of spatial location by place
cells in rat hippocampus. The spike phase relative to theta oscillations in
the EEG encodes additional information about the location of the
rat~\cite{OKeefe1993}. Another example for phase coding was reported in the
visual system. \cite{Neuenschwander1996} have found that spatially extended
light stimuli can synchronize the ongoing oscillatory activities in retinal
ganglion cells and that this synchronization is transmitted by LGN
neurons. Information encoded in phases of ongoing oscillations in synaptic
input could be used by neurons in various ways: it could be recoded,
transformed, or, like in the LGN cells mentioned, it could simply be preserved
in the spike train and thus be used to transmit information to the next
processing stage downstream. Here we concentrate on investigating the role of
ongoing oscillations in information transmission by asking how a neuron's
spike train can capture the phase of an oscillatory trend in its synaptic
input. The methods we will describe are general in that they examine a
necessary condition that observed ongoing oscillations can have a role in
transmitting information, independent of the coding scheme and of knowing what
information is transmitted.

Specifically, our aim is to model and measure how information in oscillation 
phases can be transmitted by a single neuron and how this transmission can 
coexist with information transmission utilizing spike rate. As will be shown 
in a model, the two information channels do not interfere and can be used to 
{\em multiplex} information if the oscillations are on a faster time scale 
than the rate modulations. Further, we introduce a method to quantify the 
additional information encoded in the second channel. If this additional 
information is insignificant, the role in information transmission can be 
ruled out. Importantly, however, a positive result hints to a functional role 
of the ongoing oscillations and encourages further investigations addressing 
what specific coding scheme is employed and what stimulus information 
is encoded. Thus, our method is a first step in assessing the functional 
significance of ongoing oscillations and it can be applied, even if it is 
unknown what kind of information is encoded.

We applied our new tools to recordings from relay cells in the lateral
geniculate nucleus (LGN) of cat from the lab of Judith Hirsch, USC.
Whole-cell recordings in vivo were used to record retinal {\em excitatory
postsynaptic potentials (EPSPs)} and the spikes they evoke in response to
naturalistic stimuli (movie sequences of 30s length). Cluster analysis of the
intracellular signal allowed us to label excitatory synaptic inputs (EPSPs) in
the intracellular signal and to separate these from spikes (for detailed
methods of experiments and clustering, see Wang et al. 2007,\nocite{Wang2007}
Koepsell et al. 2008).\nocite{Koepsell2008} Thus, the data we use in this
paper consist of pairs of spike trains for each geniculate relay cell, the
train of spikes produced by the relay cell and the train of presynaptic spikes
produced in the retinal ganglion cell(s) projecting to the relay cell. Our
method reveals additional information that is not captured in the classical
rate-coding paradigm. Remarkably, the amount of additional information is
commensurate with the rate-coded information and can in some cases even
surmount it.

The paper is outlined as follows: To make the paper self-contained, Section~2
briefly revisits point process models for spike trains. In Section~3, a model
is presented for reproducing the periodicity observed in measured neural spike
trains. In Section~4, an information theoretical approach is described to
measure information in oscillatory activity. In Section~5, we fit the
parameters of the model to reliably reproduce the properties of periodic spike
trains from cells recorded in LGN. Furthermore, we quantify the information
rates in spike trains with oscillations. Finally, in Section~6 we discuss
possible interpretations of the results from the information theoretic
analysis regarding potential computational functions of the oscillations in
retina and LGN.

\section{Point process models for spike trains}
In this section, we review the use of point processes in order to model a
sequence of action potentials or spikes. Each spike is an impulse of about one
millisecond duration with a stereotyped shape~\cite{Hodgkin1952}. Thus, we can
characterize the activity of a neuron by its {\it spike train}, the set of
time points $\{t_1, t_2,...,t_n\}$ at which the spikes occur.

\subsection{Homogeneous point processes} 
The simplest case is when spikes are described as homogeneous Poisson processes, 
that is, point processes that occur independently with a probability density or 
rate $\lambda$ that is constant in time. If the time axis is divided into small 
bins $dt$, the probability of a spike in any of the time bins is $\lambda
dt$. The observation of an inter-event interval $\tau_i=t_i-t_{i-1}$ relies on
the conjunction of a spike occurring in the time bin at $t_i$ and the
``survival'' of a preceding empty interval of length $\tau_i$. The survival
probability of the spike-free interval is given by the joint probability that
neither of its $\tau_i /dt -1$ time bins contains a spike. Thus, the
probability of an inter-spike interval is given by the product of
probabilities of $\tau_i /dt$ independent events, which can be approximated by
the {\it exponential distribution}:
\begin{equation}
p(\tau_i)dt=\lambda dt(1-\lambda dt)^{\tau_i/dt -1}\approx \lambda e^{-\lambda \tau_i} dt
\label{homexp}
\end{equation}
This approximation is valid for $\lambda dt << 1$ and $\tau_i/dt >>1$ and
becomes exact for $dt\to0$.

One can also compute the interval distribution between event pairs in a train
of independent spikes that are not directly consecutive but have k-1 spikes in
between, which leads to the {\it Gamma distribution} with shape parameter $k$:
\begin{equation}
p(\tau)dt=\lambda dt\frac{(\lambda \tau)^{k-1}}{(k-1)!}(1-\lambda dt)^{\tau/dt -k}
\approx \frac{\lambda^k \tau^{k-1}e^{-\lambda \tau}}{\Gamma(k)} dt\,,
\label{homgam}
\end{equation}
where $\Gamma(k)$ is the Gamma function. If the shape factor in the Gamma
distribution is an integer, it is an {\it Erlang distribution}. Note that the
Gamma distribution describes the interval density in a subsampled Poisson
process consisting of every $k$-th spike. Therefore, the Gamma distribution
for $k=1$ reduces to the exponential distribution. However, as $k$ becomes larger
than one, the shape of the distribution changes qualitatively: Whereas the
exponential distribution decreases monotonically with increasing interval size,
the Gamma distribution for $k >1$ is maximal for a certain intermediate
interval length.

It has been known for some time that the Gamma distribution matches
inter-spike-interval (ISI) distributions of real spike trains in response to
stationary stimuli much better than a exponential
distribution~\cite{Kuffler1957,Perkel1967}. Further, it has been shown that
the Erlang probability density describes the ISI distribution of a non-leaky
stochastic integrate and fire model with a fixed threshold that is driven by
excitatory Poisson inputs with constant rate~\cite{Tuckwell1988}. However, one
should add that the absence of short intervals in real spike trains is caused
by the neuronal refractory period, a short-lived history effect that prevents
a spike from being generated briefly after another spike. For the sake of
simplicity we will resort in the reminder to the use of Gamma distributions
although explicit modeling of the refractory period should yield subtle
differences. For example, the Gamma process eliminates leading and trailing
spikes of short intervals with equal probability whereas modeling the
refractory period should preferentially remove, or suppress, the trailing
spike.

\subsection{Inhomogeneous Gamma process} 
Whereas the spike rate of the homogeneous point process is constant, the
firing rate of a neuron is generally a function of time, $\lambda(t)$. Changes
in firing rate reflect changes in the stimulus, as well as other time
dependent influences on the neuron. Rate changes that are systematic and
reproducible are informative for understanding what the neuronal activity is
encoding. Equation (\ref{homgam}) can be generalized to describe the
probability density of the inhomogeneous Gamma process~\cite{Barbieri2001}
\begin{equation}
p_t(\tau)=\frac{k \lambda(t+\tau)}{\Gamma(k)}\left( k\int_{t}^{t+\tau} \lambda(u)du \right)^{k-1}
\exp\left(-k \int_{t}^{t+\tau} \lambda(u) du\right)\,,
\label{inhomgam}
\end{equation}
where $p_t(\tau)$ is the probability of a spike interval $\tau$ given that the
last spike was at time $t$. To model an actual spike train, the function
$\lambda(t)$ has to be estimated. In principle, this can be done by low-pass
filtering of the raw spike train. However, on single spike trains this
estimate is noisy and thus it is hard to separate systematic rate changes from
estimation errors and noise fluctuations. In the next section we describe
experimental designs and methods for using equation (\ref{inhomgam}) to model
spike trains that are not only influenced by a time-dependent stimulus but also 
by an oscillatory trend.

\section{Modeling systematic structure in spike trains}
Here, we model two different types of systematic structure in spike trains,
stimulus-locked rate changes and autocorrelative structure such as
periodicities in the firing rate. Both structures can be superimposed in
inhomogeneous spike trains, and both can potentially contribute to the neural
code.

\subsection{Modeling spike trains with stimulus-dependent rate modulation}

\label{sec:stilo}

To track systematic rate changes $\lambda_s(t)$ that are locked to the
stimulus $s(t)$, one typically chooses experimental designs with repetitions
of the same stimulus.  Then one can average the spike train over stimulus
repeats to form the {\it post-stimulus time histogram (PSTH)}, $r(t)$. Using
optimization techniques, such as adaptive kernel
estimation~\cite{Richmond1990}, one can find the best parameters for the
low-pass filter to estimate $\lambda_s(t)$ from the PSTH $r(t)$.  For a given
number of repeats, the power of $\lambda_s(t)$ and the inter-trial variability
can be estimated as described in~\cite{Sahani2003}.

To model a spike train with stimulus-locked rate modulation one first has to
estimate the trial-averaged rate $\lambda_s(t)$ from the PSTH and the mean
rate $\bar{\lambda}_s = 1/T\int_{0}^T\lambda_s(u)du$. Next, the time is
rescaled with
\begin{equation}
t'=\bar{\lambda}_s^{-1} \int_{0}^{t}\lambda_s(u)du
\label{trescale}
\end{equation}
in order to obtain a constant rate $\lambda=1$~\cite{Brown2002}. Finally, the
rescaled distribution of inter-spike intervals from the experiment can be
fitted by a homogeneous Gamma distribution~(\ref{homgam}) with fixed rate
$\lambda=1$ \cite{Kuffler1957}. The shape parameter $k$ is determined from the
moments (mean $\bar{\tau}$ and variance $\sigma_{\tau}^2$) of the empirical
rescaled ISI distribution as: $k=\bar{\tau}/\sigma_{\tau}^2$, see
~\cite{Barlow1957,Barbieri2001}.

\subsection{Modeling oscillatory rate modulation}
Stimulus-locked oscillatory structure is preserved by trial averaging and
therefore directly reflected in the PSTH. Thus, the oscillatory spike rate can
be obtained from trial averaging and used to model the spike train, just as
described in the previous subsection. However, neural oscillations that are
not stimulus-locked have to be modeled differently since they average out
across multiple stimulus repetitions and are therefore not captured in the
PSTH.  In the reminder we describe how to detect and quantify such
oscillations and ultimately, how to model the spike trains.

\subsubsection{Detecting and quantifying oscillations}
\label{sec:os}
Neural oscillations have been assessed in the time domain based on
autocorrelograms and inter-spike-interval (ISI) histograms. If stable
oscillations are present, a modulated autocorrelogram reflects the periodic
structure and also the ISI histogram becomes
multi-peaked~\cite{Heiss1966,Ogawa1966,Munemori1984,Castelo-Branco1998}.
Alternatively, spectral methods have been developed. Because spectral power
estimates based on the plain Fourier transform of spike trains are quite
noisy, multi-taper methods have been applied to improve the accuracy of
detecting and characterizing peaks in the power spectrum~\cite{Jarvis2001}.
Recently, the {\it oscillation score (OS)} has been proposed which exploits
structure in time and frequency domain~\cite{Muresan2008}. To compute the
oscillation score, the autocorrelogram is first filtered and the central peak
is removed (since it is not indicative of oscillatory structure).  Next, the
result is Fourier transformed and the peak frequency is detected in the
spectrum.  Finally, the oscillation score is defined as the ratio between peak
height and baseline in the spectrum.  The oscillation score is a reliable
indicator for oscillatory trends because it is insensitive to confounding
factors, such as refractory effects and bursts.

% *******************************************
% Figure 1.1 (ISI and ACOR) about here
% *******************************************
\begin{figure}[ht]
\begin{center}
\includegraphics[width=.95\linewidth]{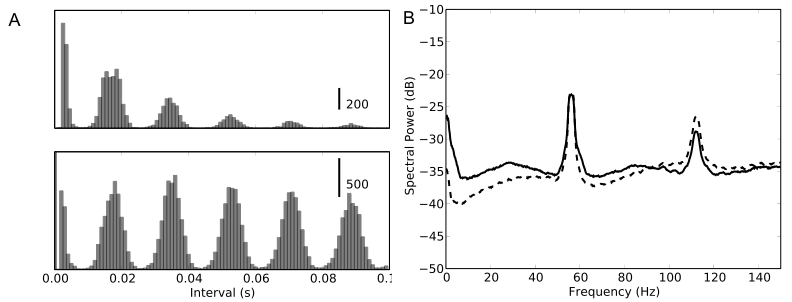} 
\end{center}
\caption{{\em Spike interval distribution, autocorrelation histogram and spike
    power spectrum for an oscillatory LGN cell.}
  \textbf{A}~Inter-spike-interval (ISI) distribution (top panel) and
  autocorrelation histogram (bottom panel) of one example LGN cell ($f =
  56.6$~Hz, $OS=39.3$). \textbf{B}~Spectral power for spikes (continuous line)
  and EPSPs (dashed line) of an example cell (same cell as in A)}
\label{fig:1.1}
\end{figure}

Fig.~\ref{fig:1.1} displays autocorrelograms and ISI histograms for relay
cells of the lateral geniculate nucleus of the cat during visual stimulation
with natural stimuli. As one can see, for this cell the oscillations are
prominent according to either of the described criteria. The timings of
synaptic inputs (EPSPs) of this cell are shown in Fig.~\ref{fig:1.2}A
together with the spikes they evoke (Fig.~\ref{fig:1.2}B). The cell belongs
to a subset of relay cells in the LGN with oscillatory spike trains that
otherwise behave quite regularly, having the usual center-surround receptive
fields and refractory periods of about one millisecond (see upper panels), see
\cite{Koepsell2008} for more details.

\subsubsection{Estimating oscillation amplitude and phase}
\label{sec:osciampphase}

\begin{figure}[ht]
\begin{center}
\includegraphics[width=.95\linewidth]{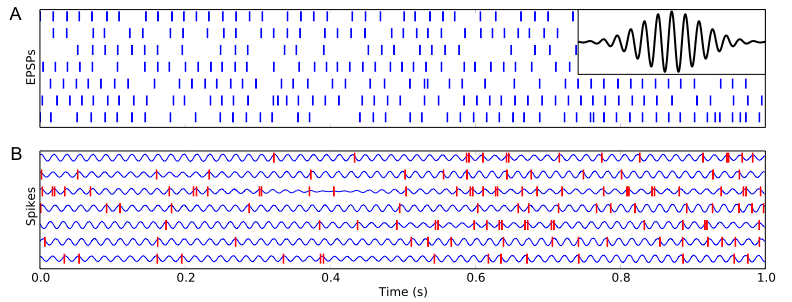} 
\end{center}
\caption{{\em Timing of retinogeniculate EPSPs and thalamic spikes recorded
    intracellularly from a single relay cell during the presentation of
    natural movies.}  \textbf{A}~Rasters of timings of EPSPs for 7 repeats
  of a natural movie clip. Inset: Morlet wavelet used to compute the analytical
  signal in B. \textbf{B}~Real part of the analytical signal computed by
  band-pass filtering of the EPSP train (blue curve) and timings of LGN spikes
  (red markers).}
\label{fig:1.2}
\end{figure}

To model the periodic structure of the spike train induced by afferent inputs
to the neuron, the instantaneous oscillation phase in the input has to be
estimated for each point in time (see Fig.~\ref{fig:1.2}). For LGN cells it
known that the synaptic inputs can exhibit periodicity originating from the
retina~\cite{Heiss1966,Ogawa1966,Rodieck1967,Castelo-Branco1998}. The LGN
recordings we analyze here are in vivo whole-cell recordings and we can
extract the train of excitatory postsynaptic potentials (EPSPs) --the input
spike train-- from the measured membrane potential.

To determine the instantaneous phase of the input oscillation of an LGN cell we 
compute the complex analytic signal of the input spike train (EPSPs)
\begin{equation}
A(t) = A_0(t) exp(i\phi(t))
\label{anasig}
\end{equation}
by convolving the EPSP train with a complex Morlet wavelet (inset of
Fig.~\ref{fig:1.2}A)
\begin{equation}
w(t,f)= C e^{2\pi ift} e^{-t^2/2 \sigma_t^2}
\label{wavelet}
\end{equation}
centered at a frequency $f$ with temporal width $\sigma_t$ and normalization
factor C. The amplitude $A_0$ of the analytic signal corresponds to the local
power in the frequency band centered at $f$ with bandwidth $\sigma_f =
1/(2\pi\sigma_t)$. 

% *******************************************
% Figure 1.2 (spike train) about here
% *******************************************

\begin{figure}[ht]
\begin{center}
\includegraphics[width=.95\linewidth]{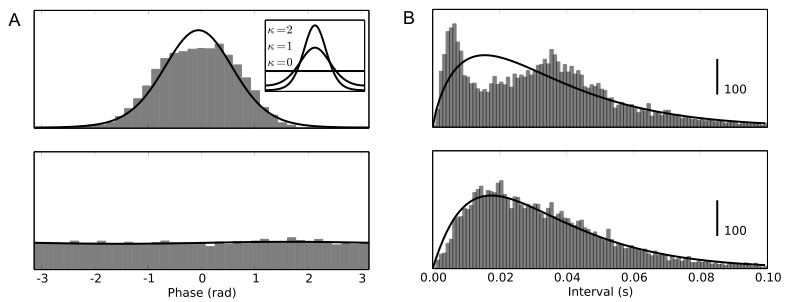} 
\end{center}
\caption{{\em Phase distribution of spikes and ISI distributions after time
    rescaling.}  \textbf{A}~Phase distribution of spikes with respect to
  oscillation extracted from the EPSP train (top panel) and Shift predictor -
  phase distribution of spikes with respect to oscillation extracted from EPSP
  train from other trial (bottom panel). \textbf{B}~Inter-spike-interval (ISI)
  distribution rescaled by stimulus-influenced rate $\lambda_s$ (top panel)
  and ISI distribution rescaled by the modulated rate used in the QPG
  model~(\ref{lambdamulti}) (bottom panel).}
\label{fig:2}
\end{figure}

The angle $\phi(t)$ of the analytic signal defines the instantaneous phase of
the input oscillation. To assess the influence of the input oscillation on the
LGN spike train, we measure how the spikes are distributed over the phase of
the input oscillation (see Fig.~\ref{fig:1.2}B). The top panel in
Fig.~\ref{fig:2}A shows that the resulting phase histogram is peaked, that is,
the spikes are more likely to occur at a certain phase in the input
oscillation. If one uses the input phase from another trial to form the phase
histogram of spikes (shift predictor), the resulting histogram shown in the
bottom panel of Fig.~\ref{fig:2}A is flat, indicating that the input
oscillation is not locked to the stimulus. To determine the optimal frequency
for this analysis we used the frequency determined from the oscillation score
measurement (see Section~\ref{sec:os}). To quantify the spike locking to the
input oscillation, the phase histogram is fitted with a von Mises (or cyclic
Gaussian) distribution
\begin{equation}
M(\phi| \kappa, \mu) = e^{\kappa cos(\phi - \mu)}/(2 \pi I_0(\kappa))
\label{vonmises}
\end{equation}
The mean phase $\mu$ is computed from circular mean of the spike phase distribution
\begin{equation}
  \langle e^{i \phi}\rangle = \frac{1}{N}\sum_{n=1}^N e^{i \phi(t_n)} = r e^{i\mu}
\label{fit-mu}
\end{equation}
The concentration parameter $\kappa$ is obtained by numerical solution of the
equation
\begin{equation}
I_1(\kappa)/I_0(\kappa) = r
\label{fit-kappa}
\end{equation}
where $I_0$ and $I_1$ are the modified Bessel functions of zeroth and first
order. The concentration parameter $\kappa$ is a measure of phase locking; the
phase distribution becomes uniform for $\kappa\to0$ and approaches a Gaussian
distribution with standard deviation $\sigma=1/\kappa$ for large $\kappa$, see
inset in Fig.~\ref{fig:2}A. The phase concentration is often characterized by
the circular variance
\begin{equation}
\textrm{var}(\phi) := \langle |e^{i \phi}|^2\rangle-|\langle e^{i \phi} \rangle|^2 
\label{circ-var}
\end{equation}
which assumes values between zero and one. The circular variance is related to
the concentration parameter $\kappa$ by $\textrm{var}(\phi) =
1-I_1(\kappa)^2/I_0(\kappa)^2$. The von Mises Distribution
$M(\phi|\kappa,\mu)$ is the maximum entropy distribution for a given circular
mean and variance.

% *******************************************
% Figure 2 (Phase histo and dewarped ISIs) about here
% *******************************************

\subsubsection{The quasi-periodic gamma model}
\label{sec:QPG}

\begin{figure}[ht]
\begin{center}
\includegraphics[width=.95\linewidth]{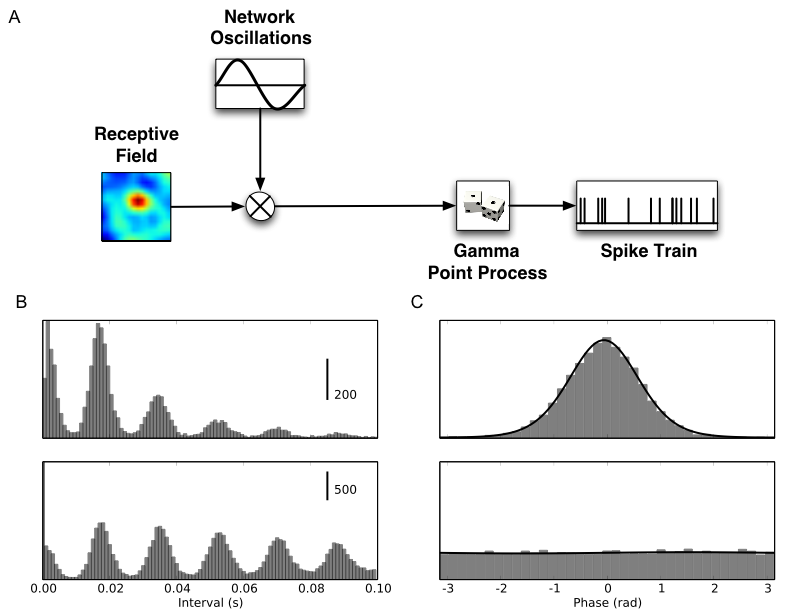} 
\end{center}
\caption{{\em Quasi-periodic Gamma (QPG) model and simulated data.}
  \textbf{A}~Schematics of quasi-periodic gamma model. \textbf{B}~ISI
  distribution (top panel) and autocorrelation histogram (bottom
  panel). \textbf{C}~Phase histogram of spikes with respect to oscillatory
  trend. The simulations of panel B and C use parameters fitted to the cell
  shown in Fig.~\ref{fig:1.1} and Fig.~\ref{fig:2}.}
\label{fig:4}
\end{figure}

To understand the effect of the combination of stimulus-dependent rate
modulation and the influence of input oscillations that are not locked to the
stimulus, we devised a simple model to include both effects, the
quasi-periodic gamma (QPG) model (Fig.~\ref{fig:4}A). It describes spike
generation by an inhomogeneous Gamma process (\ref{inhomgam}) with a factorial
instantaneous rate $\lambda(t)$ given by the product
\begin{equation}
\lambda(t) = 2\pi \lambda_s(t) M(\phi(t)| \kappa, \mu)\,.
\label{lambdamulti}
\end{equation}
The first factor $\lambda_s(t)$ is the stimulus-locked rate determined as
described in Section~\ref{sec:stilo}. The second factor is a von Mises
distribution $M(\phi(t)| \kappa, \mu)$ describing the periodic modulation that
can be fit to data as described in Section~\ref{sec:osciampphase}. In order to
get independent estimates of stimulus-locked and periodic modulations, it is
important that the oscillations have a higher frequency than the frequency
content of the stimulus-locked modulation. For LGN cells in cat these two
frequency bands are well separated, since the stimulus signal rolls off at
frequencies well below the ones of the observed oscillations in the gamma
frequency band (40-80~Hz). In the following, we use the phase of a random
band-pass signal with frequency $f\pm \sigma_f$. In total, the QPG model has
the five free parameters: $k, \kappa, \mu, f, \sigma_f$. In general, however,
the instantaneous phase $\phi(t)$ of the periodic activity may be a function
of the stimulus. The relations between the described QPG model and previous
models will be considered in the discussion section.

% *******************************************
% Figure 4 (Model, modeled ISI histograms) about here
% *******************************************

\section{Information in oscillatory spike trains}
\label{sec:info}

In this section we describe how the information content in oscillatory spike
trains can be estimated. Various methods have been developed for estimating
information rates of neural responses, e.g. \cite{Eckhorn1975, Rieke1999}, for
an overview, see Ref.~\cite{Borst1999}. Most of these methods depend on
certain properties of the statistics of signal and noise in the stimulus as
well as in the neural response, for example many models assume
Gaussianity. Here, we apply the direct method which is applicable to signals
with arbitrary statistics to the QPG model of Section~\ref{sec:QPG}.

\subsection{Direct method}

The direct method~\cite{Rieke1999,Brenner2000} estimates the mutual
information between an event $E$ in the neural response and a time-varying
stimulus $s(t)$. If the information is conveyed by single occurrences of event
$E$ (and not temporal patterns), the information per event is given by
\begin{equation}
I(E;s)=S[E]-S[E|s] \;\;\mbox{bit/event}\,.
\label{mutinfo}
\end{equation}
If the distribution of event $E$ is uniform during the trial $t \in (0,T)$,
that is, $p(t)= 1/T$, the unconditional entropy is simply $S[E] = \log_2
T$. $S[E|s]$ is the conditional entropy of the event $E$, given a time-varying
stimulus $s(t)$:
\begin{eqnarray}
S[E|s] &=& \sum_{s \in \{s\}} p(s) \int_0^T dt\; p(t|s) \log_2\left(p(t|s)\right)\nonumber\\
       &=& \int_0^T dt\; p(t|s) \log_2\left(p(t|s)\right)
\label{condinfo}
\end{eqnarray}
If the stimulus is rich enough, the ensemble average over stimuli and the time
average over the stimulus are equivalent~\cite{Brenner2000} and thus the
former can be omitted in equation~(\ref{condinfo}).

\begin{figure}[ht]
\begin{center}
\includegraphics[width=.95\linewidth]{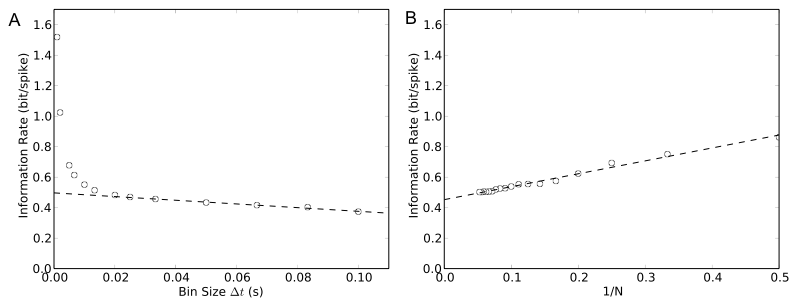} 
\end{center}
\caption{{\em Information rates for experimental data (natural stimulus with
    20 repeats).}  \textbf{A}~Information rate as a function of the size of
  the time bin (circles). Linear extrapolation to zero bin width yields 0.50
  bit/spike (dashed line). \textbf{B}~Information rate as a function of the
  inverse number of repeats (circles). Linear extrapolation to infinite number
  of trials yields 0.45 bit/spike (dashed line). }
\label{fig:3}
\end{figure}

The conditional distribution of spikes can be estimated empirically by
recording neural responses to repeats of the stimulus $s(t)$ and forming
the PSTH $r(t)$ (see Section~\ref{sec:stilo}). The conditional distribution 
is then given as $p(\mbox{spike at } t|s) = r(t)/(T \bar{r})$. Thus, each 
spike transmits the information~\cite{Brenner2000}:
\begin{equation}
I(\mbox{spike};s)=\frac{1}{T}\int_{0}^{T} dt\; 
\frac{r(t)}{\bar{r}}\log_2\left(\frac{r(t)}{\bar{r}}\right)\mbox{bit/spike}
\label{infodirect}
\end{equation}
The accuracy of the information rate estimated for finite data depends on the
bin width $\Delta t$ used to compute the integral in equation
(\ref{infodirect}). The estimate converges to the true entropy only
asymptotically (limit of zero bin width and infinite number of trials). More
specifically, narrow bin sizes with finite data lead to a pronounced
overestimation of the amount of information transmitted. The circles in
Fig.~\ref{fig:3}A show how for a data set with 20 trials the information is
overestimated as the bin size decreases. The estimate is improved, however, by
a linear extrapolation ($\Delta t\to0$) of the values for larger bin size,
Fig.~\ref{fig:3}A, dashed line.  The resulting value is 0.50 bit/spike, see
(see~\cite{Brenner2000}).  Extrapolating this result in addition to larger
numbers of trials (N) also with a line ($\Delta t,1/N\to0$) yielded 0.46
bit/spike (Fig.~\ref{fig:3}B, dashed curve).

% *******************************************
% Figure 3 (Info direct data) about here
% *******************************************

\subsection{The multiconditional direct method}
\label{sec:multicond}

In the following, we apply the direct method to the case of oscillations that
are not locked to the stimulus. If an oscillation is present, the spike train
contains information not only about the stimulus signal $s(t)$ but also about
the phase $\phi(t)$ of the oscillatory trend. To quantify the contribution of
the oscillation to the information rate, the two cases described in
Section~\ref{intro} have to be treated differently; stimulus-locked
oscillations and oscillations that are not locked to the
stimulus. Stimulus-locked oscillations are conveyed in the PSTH and therefore
equation~(\ref{infodirect}) can be used directly to estimate the mutual
information in single spikes exerted by the stimulus and the oscillatory
trend.

Oscillatory trends that are not locked to the stimulus are averaged out in the
PSTH. Therefore the additional information about the oscillation cannot be
measured by equation (\ref{infodirect}) \footnote{But note that equation
  (\ref{infodirect}) can still be used to estimate the information in
  stimulus-locked rate modulation.}. Here, we describe the multiconditional
direct method that can measure information in oscillatory activity, even if
not locked to the stimulus. We quantify the information
$I(\mbox{spike};s,\phi)$ that a spike conveys about both the stimulus $s(t)$
and the phase $\phi(t)$ of an oscillatory trend. Note that this is different
from quantifying the stimulus information conveyed by a spike in a phase
coding scheme, which would be $I(\mbox{spike},\phi;s)$ (see Discussion).

The periodic trend function can be described by an instantaneous phase
$\phi(t)$ which increases monotonically with wrap-around condition $\phi =
\phi+2\pi$. The phase $\phi(t)$ is a quasi-periodic function in time with
period $P(t)$: $\phi(t)=\phi(t+P(t))$. The probability of a spike occuring at
time $t$ now depends on both influences, the stimulus and the oscillatory
trend function. At any moment in time $t$, the joint influence is fully
described by the stimulus $s(t)$ and the current phase of the oscillatory
trend $\phi(t)$. The conditional entropy that includes both influences can
then be written:
\begin{eqnarray}
S\big[p(\mbox{spike at } t|s, \phi)\big]
&=& \sum_{s, \phi} p(s,\phi) \int_0^T dt\; p(t|s,\phi) \log_2\left(p(t|s, \phi)\right)\nonumber\\
&=&  \sum_{s} p(s) \sum_{\phi} p(\phi|s)\int_0^T dt\; p(t|s,\phi) \log_2\left(p(t|s,\phi)\right)\nonumber\\
&=& \int_0^{P} d\phi\int_0^T dt\; p(t|s, \phi) \log_2\left(p(t|s,\phi)\right)
\label{condentroext}
\end{eqnarray}
Here the same argument applies for dropping the ensemble average over the
stimuli as in equation (\ref{condinfo}). In addition, we use the fact that the
oscillations are not stimulus locked and therefore the instantaneous phase
does not depend on the stimulus and all phases occur equally often:
$p(\phi|s)=1/(2\pi)$.

As before, the conditional distribution $p(t|s,\phi)$ can be acquired
empirically by recording neural responses during multiple repetitions of the
stimulus $s(t)$. The response varies not only with the stimulus but also with
the instantaneous oscillation phase $\phi(t)$. To capture both dependencies,
an extended response histogram has to be collected. The stimulus can be
parameterized by the relative stimulus time, just as in the standard PSTH. But
because the oscillations are not stimulus-locked, the histogram requires a
second dimension to represent the instantaneous phase $\phi(t)$. The extended
response histogram $r(t,\phi)$ contains the response binned by relative
stimulus time $t\in[0,T]$ and instantaneous phase $\phi\in[0,2\pi]$. With
proper normalization the conditional probability is given by: $p(t|s,\phi) =
r(t,\phi)/(T\bar{r})$. The unconditional probability of a spike at time $t$ is
$p(t)= 1/T$ with entropy $S[p(t)]\,=\,\log_2T$ as before.

Using equation (\ref{mutinfo}) the {\it multiconditional direct method} can be
formulated: The information per spike is given as:
\begin{equation}
I(\mbox{spike};s,\phi)=\frac{1}{2\pi T}\int_0^{2\pi} d\phi \int_{0}^{T} dt\;
\frac{r(t, \phi)}{\bar{r}}\log_2\left(\frac{r(t, \phi)}{\bar{r}}\right)\mbox{bit/spike}
\label{infodirectext}
\end{equation}
One might ask, what the maximal amount of information is that could be encoded
in the spike train this way. If the stimulus-locked rate and the periodic
oscillations would encode information independently, which would be optimal,
their respective contribution to the right hand side of formula
(\ref{infodirectext}) would be additive. The contributions from the
oscillations is bounded by the negative entropy of the von Mises distribution
\begin{equation}
\label{eq:misent}
S\big[M(\phi|\kappa,\mu)\big] = 
  \log_2\left(\frac{\exp(\kappa I_1(\kappa)/I_0(\kappa))}{2\pi I_0(\kappa)}\right)\,.
\end{equation}
Thus, the upper bound of the total information rate of single spikes in oscillatory
spike trains is given by
\begin{equation}
\label{eq:bound_dirext}
I(\mbox{spike};s,\phi) \le I(\mbox{spike};s) - S\big[M(\phi|\kappa,\mu)\big].
\end{equation}
with the first term on the RHS obtained from (\ref{infodirect}) and the second
term from (\ref{eq:misent}). 
A deviation from this upper bound indicates mutual information between the
phase and the rate signal which might be utilized for redundant encoding of
information (see Discussion).

\subsubsection{The phase de-jittering method}
\label{sec:de-jitter}
An alternative option to determine the information is a de-jittering method,
proposed in Ref.~\cite{Koepsell2008}, in order to measure the information
carried by single spikes in trials with different values for the instantaneous
phase $\phi(t)$ of the oscillation in the inputs. Here, we shift the spikes by
an amount corresponding to their oscillation phase $-\phi(t)/(2\pi f)$ in
order to correct for the effect of different state of the input oscillation
and subsequently measure the single-spike information using the direct method
(\ref{infodirect}). This amounts to an oscillation-dependent rescaling of the
time axis and effectively {\em de-jitters} the spike train.

This method relies on the fact that the stimulus-locked spike rate does not
change on the time scale of the oscillations and therefore is not affected by
the de-jittering. Even though this is only an approximation, the de-jittering
method gives comparable estimates to the multiconditional direct
method~(\ref{infodirectext}) and it has the advantage that it requires much
less data to be feasible.

Note that the described method of de-jittering differs from those that use the 
stimulus~\cite{Aldworth2005} or the spike train itself~\cite{Richmond1990}.

\section{Model fitting and simulation experiments}
\label{sec:model}

In this section, we describe how to fit the QPG model to measured data and
apply it to intracellular recordings of thalamic relay cells in cat LGN. We
then use the QPG model to generate different types of surrogate data for 
testing and comparing the information theoretic measures described in 
Section~\ref{sec:info}. 

\subsection{Fitting the QPG model to LGN recordings}

We fitted the five free parameters ($k, \kappa, \mu, f, \sigma_f$) of the
model to match the data of the cell shown in
Fig.~\ref{fig:1.1}A. Specifically, frequency and bandwidth of the oscillatory
component was fitted to the oscillations measured in the input (train of
EPSPs), the concentration and mean of the spike phases was fitted to the
experimental phase histograms and the shape factor of the Gamma distribution
was optimized to match the output spike train.  The center frequency of the
oscillation was set to the frequency determined by the oscillation score
method (see~\cite{Muresan2008} and Section~\ref{sec:os}) and for the bandwidth
we chose $\sigma_f=2$Hz, corresponding to the width of the peak in the power
spectrum (Fig.~\ref{fig:1.1}B) and a temporal width of $\sigma_t=80$ms. To
generate the oscillatory component, a bandpass filter with these parameters
was applied to a white-noise signal throughout the entire recording length,
irrespective of stimulus onsets. Thus, the resulting phase distribution across
trials was flat, as in the experimental data (Fig.~\ref{fig:2}B). The
concentration parameter $\kappa$ and the mean phase $\mu$ were determined by
fitting the von Mises distribution to the phase distribution of the spikes
using equations (\ref{fit-mu}) and (\ref{fit-kappa}). Finally, the shape
parameter $k$ of the Gamma process was determined by using time
rescaling~\cite{Brown2002} as described in Section~\ref{sec:stilo} (see
Fig.~\ref{fig:2}B).

The QPG model reproduces the characteristics of the real data quite well. The
simulated ISI histogram and autocorrelogram in Fig.~\ref{fig:4}B is very
similar to the histograms from the data in Fig.~\ref{fig:1.1}B and also the
histograms of spike phases are similar: simulated data shown in
Fig.~\ref{fig:4}C and real data in Fig.~\ref{fig:2}A. Another indication that
the QPG model captures the measured spike train very well is the good match
between the time-rescaled ISI histogram using equation (\ref{lambdamulti}) and
the Gamma distribution, compare histogram and solid line in Fig.~\ref{fig:2}B
(bottom panel). As a control we rescaled the ISI histogram just by
$\lambda_s$, ignoring the oscillations. The resulting histogram is clearly
not well described by a Gamma distribution, Fig.~\ref{fig:2}B (top panel).

\subsection{Estimation of Information rates}

% *******************************************
% Figure 5.1 (Info direct simulation) about here
% *******************************************
\begin{figure}[ht]
\begin{center}
\includegraphics[width=.95\linewidth]{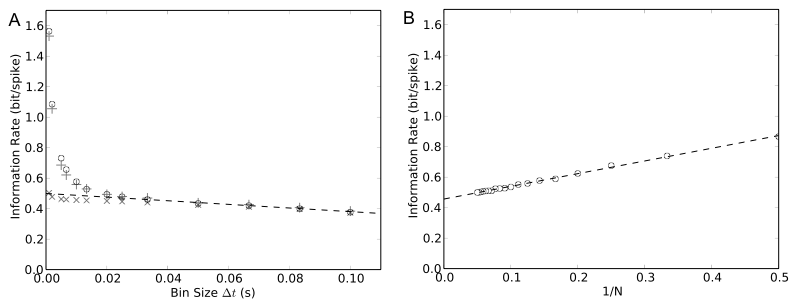} 
\end{center}
\caption{{\em Information rates for simulated data using the direct method.}
  \textbf{A}~Information rate using equation (\ref{infodirect}) on 20 trials
  as a function of the size of the time bin (circles). Linear extrapolation to
  zero bin width yields 0.50 bit/spike (dashed line). Information estimates for
  simulated data without modulation gives identical results
  (crosses). Estimates diverge for bin sizes below about $30 ms$ due to
  limited data. Information estimates using 500 trials give reliable results
  for smaller bin sizes (crosses). \textbf{B}~Information rate using equation
  (\ref{infodirect}) on 20 trials as a function of the inverse number of
  repeats (circles). Linear extrapolation to infinite number of trials yields
  0.46 bit/spike (dashed line).}
\label{fig:5.1}
\end{figure}

The fitted QPG model was used to create three different types of simulated
data. The {\it surrogate data} reflected the entire structure of the fitted
experimental data, as well as the QPG model permitted. The spike trains in the
surrogate data exhibited stimulus-locked modulation and an oscillatory trend
that was not locked to the stimulus. Since it is very difficult to conduct in
vivo recordings over long periods of time, the number of stimulus repetitions
available in the real data is very limited, in our case to 20 trials. The
surrogate data can be made arbitrarily large and thus they allow to study
effects of data limitation. Further, surrogate data allow to estimate
information measures even though the original data set would be much too small
to achieve a reliable estimation.

For the purpose of comparison we also generated data sets with the QPG model
that had systematic differences from the real data. The second type of
simulated data used all the fit parameters reflecting the stimulus modulation
and the oscillation strength of the experimental data. However, unlike in the
data the oscillatory trend in the model was aligned over repeats, that is,
this data set contained {\it stimulus-locked oscillations}.  The third type of
simulated data reflected only the stimulus modulation observed in the data but
{\it no oscillatory structure}.

First we asked how the results with the standard direct method depends on the
amount of data. We computed information rates with equation~(\ref{infodirect})
for 20 trials (same number of trials as in the data) and for 500 trials from
the surrogate data. Fig.~\ref{fig:5.1} shows the information rates for the
surrogate data in the same format as Fig.~\ref{fig:3}. The results for 20
trials (circles in in Fig.~\ref{fig:5.1}A) exhibit a similarly strong bias for
bin sizes smaller than $30 ms$ as is also observed for the experimental data
(Fig.~\ref{fig:3}A).  Using the linear extrapolation from larger bin sizes to
estimate the information rate as in Fig.~\ref{fig:3}, the results for the
surrogate data are very similar to the results on the real data in
Fig.~\ref{fig:3}: 0.50 bit/spike for extrapolation to small time bins ($\Delta
t\to0$) Fig.~\ref{fig:5.1}A; and 0.45 bit/spike for extrapolation to small
time bins and large numbers of trials ($\Delta t,1/N\to0$),
Fig.~\ref{fig:5.1}B. In addition, the surrogate data allow to verify the
validity of the linear extrapolation. The information estimates for 500 trials
(crosses in Fig.~\ref{fig:5.1}A) lie slightly lower than the ones for 20
trials, as expected, but follow the linear trend to much smaller bin sizes.

% *******************************************
% Figure 5.2 (Info direct simulation) about here
% *******************************************
\begin{figure}[ht]
\begin{center}
\includegraphics[width=.95\linewidth]{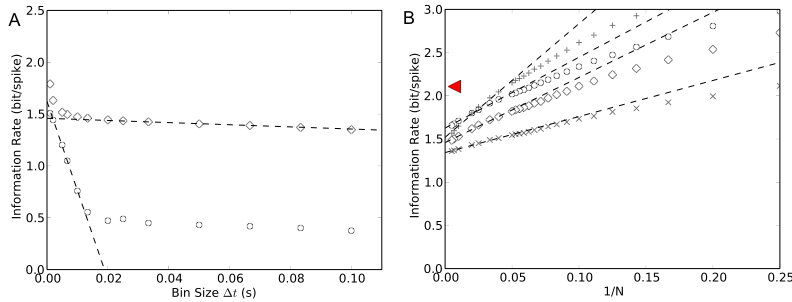} 
\end{center}
\caption{{\em Information rates for simulated data using the multiconditional
    direct method.}  \textbf{A}~Circles mark the information rate on 500
  trials as a function of the size of the time bin for stimulus-locked
  oscillations computed with equation (\ref{infodirect}). Note the additional
  information compared to A below $30 ms$ bin size. Diamonds mark the
  information rate for non-stimulus-locked oscillations using equation
  (\ref{infodirectext}). \textbf{B}~Information rate using equation
  (\ref{infodirectext}) as a function of the inverse number of trials. Circles
  mark the rates obtained with equation (\ref{infodirect}) for stimulus-locked
  oscillations. Other symbols denote results obtained with
  (\ref{infodirectext}) for non-stimulus-locked oscillations for different
  numbers of phase bins: 5 phase bins (crosses), 10 phase bins (diamonds), and
  20 phase bins (pluses). Dashed lines are the linear extrapolations. Both
  methods give comparable results around 1.5~bit/spike. The red triangle marks
  the maximal achievable amount of information rate of 2.13~bit/s computed by
  assuming independence of oscillatory and rate-encoded information, see
  equation~\ref{eq:bound_dirext}.}
\label{fig:5.2}
\end{figure}

Next, we used equation~(\ref{infodirectext}) on the surrogate data to assess
the contribution of the not stimulus-locked oscillations to the information
rate of the spike trains generated by the QPG model. For comparison, we
studied also the effect of stimulus-locked oscillations by applying the
standard formula for direct information (\ref{infodirect}) to the simulated
data with the oscillations aligned across trials. In both cases we used 500
simulated trials. Fig.~\ref{fig:5.2} depicts the resulting information
rates. For the stimulus-locked oscillations (circles) the additional
information due to the oscillations is visible only at small bin sizes in
Fig.~\ref{fig:5.2}A and therefore can easily be overlooked with limited
data. In contrast, the result of non-stimulus-locked oscillations (diamonds)
can be extrapolated from larger bin sizes. But note that the information rate
of not stimulus-locked oscillations relies on an extended response histogram
(in time and phase) and therefore the required amount of data is not smaller
than for stimulus-locked oscillations. Therefore, both methods we use in
Fig.~\ref{fig:5.2} to estimate the information rates rely on boosting the
amount of data with the QPG model, the extrapolations to zero bin size could
not have been done directly on the experimental data with only 20 repeats. In
the right panel (Fig.~\ref{fig:5.2}B) the obtained estimates for the
information rates are extrapolated to infinite number of trials using
different numbers of phase bins in the extended response histogram. Three
observations with the resulting asymptotic information rates should be
emphasized. First, oscillations contribute a significant amount of
information. They add about 1 bit/spike for this cell, more than twice the
information contained in stimulus-locked rate modulations alone
(cf. Fig.~\ref{fig:5.1}A/B). Second, the information rates have a similar
value, around 1.5 bit/spike, whether or not the oscillations are locked to the
stimulus. Note that the rates for not stimulus-locked oscillations converge to
1.5 bit/spike for 10 or more phase bins. Third, the measured information rates
reach about 70\% of the value one would expect, if stimulus-locked and
oscillatory modulation were entirely independent. The upper bound
(\ref{eq:bound_dirext}) for this cell that was best fitted with $\kappa=2.44$
is $2.13$~bit/spike, as marked by the red triangle in Fig.~\ref{fig:5.2}B.

%%sollen wir gruende fuer die diskrepanz zwischen bound und estimates diskutieren?
%%im QPG model sollte ja die unabhaengigkeit erfuellt sein 

% *******************************************
% Figure 7 (Comparison with de-jittering) about here
% *******************************************
\begin{figure}[ht]
\begin{center}
\includegraphics[width=.95\linewidth]{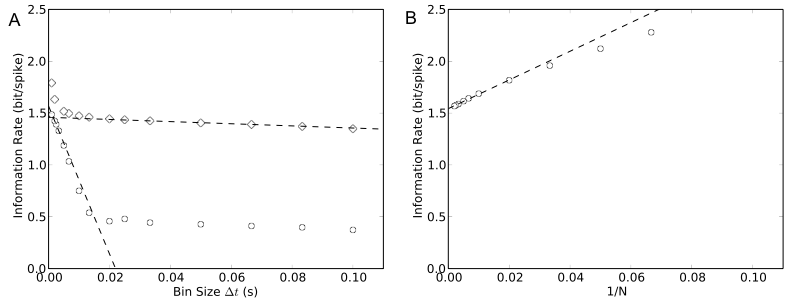} 
\end{center}
\caption{{\em Estimating Information rate using spike de-jittering method.}
  \textbf{A}~Information rate as a function of bin width for single spikes
  obtained by formula (\ref{infodirect}) after spike de-jittering (circles):
  Using a reference oscillation $\phi_0(t) = f\cdot t$, individual spikes have
  been shifted by the phase-dependent amount $\phi_0-\phi(t)/(2\pi f)$. The
  obtained information rate is comparable to the one obtained by the
  multiconditional method (diamonds). \textbf{B}~Extrapolation of
  the information estimate in A to infinite number of trials.}
\label{fig:7}
\end{figure}

% de-jittering method
We have investigated whether the de-jittering method in
Section~\ref{sec:de-jitter} gives comparable results to the multiconditional
method to estimate the information in oscillatory spike trains. The
de-jittering method has been used in Ref.~\cite{Koepsell2008} to estimate
information rates. However, since the limited amount of experimental data, the
information of the de-jittered spike train has been estimated using the
signal-to-noise ratio and the assumption of a Gaussian information
channel. Here, we de-jittered the simulated data as described in
Section~\ref{sec:de-jitter} and applied equation (\ref{infodirect}) to
determine the information rate in single spikes. Fig.~\ref{fig:7}A/B shows
that the information rate estimation using de-jittering agrees quite well with
the results obtained by the more rigorous treatment described in this
paper. This comparison confirms that the de-jittering method is a viable
method to estimate the information in oscillatory activity on limited amounts
of data without explicitly modeling the data.

\section{Discussion}
\subsection{Phase transmission versus phase coding}

The phenomenon we have focused on in this paper is phase coupling of
oscillatory neural activity between input and output of a neuron that also
responds to an externally defined stimulus. Specifically, we proposed a model
and a method to measure information transmission rates if the oscillations are
not locked to the stimulus. In the phase transmission we consider, each spike
provides information about the external stimulus and about the phase of the
input oscillation. The appropriate measure for the information rate in this
transmission process is $I(\mbox{spike};s,\phi)$.

It is important to emphasize how phase transmission is different from a
traditional spike phase code and how both schemes are related. In a spike
phase code, such as in place fields of the hippocampus, information is {\it
 encoded} by relative shifts of spikes with respect to a reference
oscillation (theta waves).  To read off the relative phases, the receiver
has also to have independent access to the oscillatory reference signal.
Accordingly, the transmitted information rate should be measured by
$I(\mbox{spike},\phi;s)$, where $\phi$ are the relative phases between
spikes and reference signal. Once oscillations are used to phase-encode
information, one can ask how neurons downstream can pick up phase structure
in the spike trains, carrying either phase encoded information or the reference
signal, and reliably reproduce (or transmit) it in their output.  This type
of phase transmission we have investigated in oscillatory LGN cells.

Two recent studies provide evidence for additional phase-coded information in
the visual system, however with respect to stimulus-locked oscillations:
Montemurro et al.~\cite{Montemurro2008} have shown that the spike phase
relative to slow ($<12$~Hz) oscillations in the local field potential of V1
which are locked to the stimulus carries additional stimulus
information. Masse and Cook~\cite{Masse2008} have shown that the spike phase
in MT relative to oscillations which are driven by the stimulus frame refresh
carries additional information. It has been an open question whether spike
phases relative to spontaneous oscillations that are not locked to the
stimulus encode additional information. For the LGN data we have provided
evidence that phase information is reliably transmitted. However, what type of
phase code is employed and what information is encoded has not been addressed
in this paper.

\subsection{Potential artifacts and sources of bias in parameter estimation}

%%%oscillation artifacts
Oscillatory artifacts due to line noise or due to frame and refresh rates in
the monitor are of concern in all studies involving oscillations in neural
responses to visual stimuli. Therefore, controls have to be conducted. To
control for line noise one can search for spectral peaks centered at the line
frequency of 60 Hz. Weak line noise artifacts were visible in some of the raw
membrane recordings, however, never in the spike or EPSP trains we
extracted. Further, artifacts in the membrane voltage were always easy to
discern from true spectral peaks by their narrow width. To control for monitor
artifacts one can use the monitor refresh signal in the phase locking analysis
described in Section~\ref{sec:osciampphase}. The resulting phase distributions
were flat. Furthermore, we observed that the oscillations persisted in
spontaneous spiking activity when the eye was closed (data not shown).

%%%%%%% noise and biases in the phase estimation procedure
Further, one has to exclude that the procedure described in
Section~\ref{sec:osciampphase} for estimating an oscillation phase from
intracellular recordings introduces ``false'' phase locking between inputs and
outputs of the LGN cells.  The phase estimation can in fact become noisy if
the event rates become low and errors in the event detection can introduce
biases. In our case, however, the event rates where quite high, above
30~spikes/second on average (see Fig.~\ref{fig:1.2}A). Furthermore, the main
imperfection in the event detection (to miss input events near spikes) would
weaken the phase locking effect and not introduce false ones.

%%%%%%% bias because oscillations are not averaged away in PSTH
If the information rate is estimated with limited data it is possible that the
oscillations are not fully averaged out in the PSTH. If the frequency bands of
oscillations and stimulus-locked modulation are separated, the standard
practice of using low-pass filtering in computing the PSTH
(Section~\ref{sec:stilo}) reduces this effect already significantly. Potential
residual oscillations in the PSTH lead to overestimation of the information in
the stimulus-dependent rate modulation.  Note that this potential bias does
not introduce an artifactual information gain due to oscillations. Actually,
it would work against finding a significant gain in information through the
oscillatory activity because the overestimated quantity is subtracted.

\subsection{Relations between our methods and earlier work}

The QPG model we devised to describe phase locking of neural responses to an
oscillatory input is related to some earlier models. In the limit of zero
bandwidth, the oscillation phase becomes $\phi(t)=f\,t$ and the QPG model
(with constant visual input) degenerates to a modulated Gamma process with a
cyclic trend~\cite{Berman1981}. Koyama and Shinomoto have proposed a model for
oscillatory spike trains that uses $1+\sigma\cos(\phi)$ instead of the von
Mises function as the factor to convey an oscillatory
trend~\cite{Koyama2005}. We chose the von Mises distribution to model phase
modulations since it is the highest entropy distribution for a given phase
concentration and is able to model arbitrary phase concentrations. In
contrast, the Koyama and Shinomoto (KS) model is limited to describing phase
concentrations below $\kappa=1.2$ corresponding to a circular variance of
$\textrm{var}(\phi)=.75$, even for maximal modulation $\sigma$. This
limitation of the KS model can be easily verified using
Eq.~(\ref{fit-mu})-(\ref{circ-var}).

We have derived the multiconditional information rate (\ref{infodirectext})
to measure the information rate through single spikes in a neuron that 
experiences two simultaneous influences, a stimulus-locked rate modulation
and a periodic rate modulation that is not locked to the stimulus.
This method is an extension of the direct method applied to single-spike 
events (\ref{infodirect}) \cite{Rieke1999,Brenner2000}.

\subsection{Functional roles of spontaneous oscillations}

If the observed oscillations are ongoing and contain no stimulus-dependent
modulation, the QPG model performs an operation similar to amplitude
modulation (AM) in a radio broadcast signal: The stimulus-dependent rate
$\lambda_s$ corresponds to the modulation signal, the oscillations correspond
to the high-frequency carrier in an AM signal. Analogous to broadcast
transmission, the stimulus signal is modulated into the high-frequency
band. However, unlike in a radio signal, the low frequency signal is still
present in the spike train (because the multiplication is between two positive
rate functions in the QPG model and therefore the resulting signal has no
symmetric amplitude envelope). Thus, the spike train contains the stimulus
signal twice, it can either be decoded by low pass filtering (usual method of
rate decoding) or by band pass filtering. If the frequencies of the stimulus
and the frequency of the ongoing oscillations are well separated, they can be
transmitted and read out independently from each other. Furthermore, the
low-pass and the band-pass signal are robust to different types of
noise. Therefore, this redundant coding scheme could be used to achieve a more
reliable readout of the stimulus signal downstream. For example, band pass
filtering could be realized by tuned intrinsic subthreshold
oscillations~\cite{Nowak1997,Hutcheon2000,Fellous2001,Tiesinga2008}.

In addition to robustness, oscillatory structure of the afferent input into
cortex could subserve various other functions. First, it can support
time-windowing in the readout for reducing crosstalk from all other neural
activity (that does not share the oscillation). Second, common oscillations in
different cells can produce synchrony between LGN spikes which has been shown
to improve the success rate to activate cortical
cells~\cite{Usrey2000,Bruno2006}, for a model, see~\cite{Kenyon2004}. Third,
it can enable oscillatory top-down attentional mechanisms to select between
specific input streams~\cite{Fries2007}.

\subsection{Roles and coding mechanisms of stimulus-influenced oscillations}

If the oscillations are stimulus-influenced, although not stimulus-locked,
they carry information about the stimulus. If this information were
overlapping or identical with the information in $\lambda_s$, this redundancy
could serve similar functions as described for stimulus-independent ongoing
oscillations. However, if the information carried in the oscillations would
represent stimulus properties not conveyed by $\lambda_s$, the oscillations
would enable a multiplexing in spike trains, that is, two different signals
could be carried by one spike train as seen e.g.\ in the olfactory
system~\cite{Stopfer2003}. The multiplex scheme could employ a number of
different encoding schemes. For example, the coherence of a spike train could
be modulated or the phase of neural oscillations or of spikes relative to
these oscillations could encode additional stimulus information
\cite{Gray1989,Neuenschwander1996,Samonds2006,Montemurro2008,Masse2008}.

The QPG model does not explicitly model how additional {\em stimulus
  properties} are encoded in the oscillatory signal. Nevertheless, we can use
the QPG model to estimate the information that can be submitted using the
oscillations as an additional information channel. Using a simple
back-of-an-envelope calculation we can estimate the information in a phase
code that synchronizes the oscillations (and thus the spikes) of two cells
\cite{Gray1989,Neuenschwander1996,Samonds2006} in a population of
neurons. Since the oscillations in the fitted model have a bandwidth of
$\sigma_f =2$Hz, phase alignments cannot be instantaneous: The maximum
relative phase adjustment necessary to synchronize two cells takes about 125
ms and is therefore within the behaviorally relevant range. Even with a
conservative information estimate of 1 bit per cell in 125 ms (corresponding
to two possible phases for say figure and ground), the encoded information
would be about 8 bit/s and therefore comparable to the rate-encoded
information of 9 bit/s (0.45~bit/spike at a typical rate of
20~spikes/s). Thus, the oscillation-based channel can encode stimulus
information in the oscillation phase that is independent from the type of
information conveyed by the rate.

If such an additional oscillation-based information channel exists in the
early visual system, the most interesting question is, of course, what
properties of the stimulus the channel conveys. So far, this question has not
be answered for the early mammal visual system. However, a recent study of the
impact of retinal oscillations on the behavior of frogs suggest the intriguing
possibility that retinal oscillations could encode nonlocal information such
as spatial or temporal stimulus context \cite{Ishikane2005}.

\section{Summary}

The paper presented new theoretical tools for studying the functional roles of
oscillatory activity in the brain. To simulate oscillatory spike trains that
are phase-locked to an oscillatory influence that is not stimulus-locked we
described the quasi-periodic gamma (QPG) model. This model generates spikes
using an inhomogeneous Gamma process modulated by the product of a
stimulus-influenced rate and a quasi-periodic von Mises distribution. The QPG
model can be fitted to oscillatory recordings in the LGN of the cat and
reproduces the main characteristics of the data, such as ISI histogram,
oscillation score and spike-phase histograms. To capture the information in
the spike train about oscillations that are not locked to the stimulus, we
propose the multiconditional direct method (\ref{infodirectext}), a
generalization of the direct method applied to single-spike
events~\cite{Rieke1999,Brenner2000}. We estimated information rates in
oscillatory LGN cells and discussed the possible consequences from our finding
that oscillations contribute significantly to the information carried in the
spike train.  However, the proposed computational methods are not confined to
visual neurons, they are general tools for investigating the transmission of
oscillatory structure in neural activity in the brain.

\vspace{1cm}

\noindent {\bf Acknowledgments:} We thank J. Hirsch, X. Wang, and
V. Vaingankar, University of Southern California, for numerous helpful
discussions and for permitting the use of their experimental data. We also
thank Tim Blanche and Charles Cadieu for valuable comments on the
manuscript. The comments of two anonymous reviewers improved the manuscript
substantially. This work has been supported by NSF grant IIS-0713657. The data
analysis and simulations were computed using IPython~\cite{Perez2007} and
NumPy/SciPy~\cite{Oliphant2007}, all figures were produced using
Matplotlib~\cite{Barrett2005}.

%\vspace{-0.3cm}
\renewcommand\refname{\normalsize References}
\small

\bibliographystyle{alpha}
\setlength{\itemsep}{0cm}
% \bibliography{oscillations}

\begin{thebibliography}{WWV{\etalchar{+}}07}

\bibitem[Adr42]{Adrian1942}
ED~Adrian.
\newblock {Olfactory reactions in the brain of the hedgehog}.
\newblock {\em The Journal of Physiology}, 100(4):459--73, 1942.

\bibitem[AMG{\etalchar{+}}05]{Aldworth2005}
Z.N. Aldworth, J.P. Miller, T.~Gedeon, G.I. Cummins, and A.G. Dimitrov.
\newblock {Dejittered Spike-Conditioned Stimulus Waveforms Yield Improved
  Estimates of Neuronal Feature Selectivity and Spike-Timing Precision of
  Sensory Interneurons}.
\newblock {\em Journal of Neuroscience}, 25(22):5323--32, 2005.

\bibitem[AV90]{Ahissar1990}
E.~Ahissar and E.~Vaadia.
\newblock {Oscillatory activity of single units in a somatosensory cortex of an
  awake monkey and their possible role in texture analysis.}
\newblock {\em Proceedings of the National Academy of Sciences of the United
  States of America}, 87(22):8935--9, 1990.

\bibitem[BBV{\etalchar{+}}02]{Brown2002}
E.N. Brown, R.~Barbieri, V.~Ventura, R.E. Kass, and L.M. Frank.
\newblock {The Time-Rescaling Theorem and Its Application to Neural Spike Train
  Data Analysis}.
\newblock {\em Neural Computation}, 14(2):325--346, 2002.

\bibitem[Ber81]{Berman1981}
M.~Berman.
\newblock {Inhomogeneous and Modulated Gamma Processes}.
\newblock {\em Biometrika}, 68(1):143--52, 1981.

\bibitem[BFK57]{Barlow1957}
HB~Barlow, R.~Fitzhugh, and SW~Kuffler.
\newblock {Change of organization in the receptive fields of the cat's retina
  during dark adaptation}.
\newblock {\em The Journal of Physiology}, 137(3):338--54, 1957.

\bibitem[BHM{\etalchar{+}}05]{Barrett2005}
P.~Barrett, J.~Hunter, JT~Miller, J.C. Hsu, and P.~Greenfield.
\newblock {matplotlib--A Portable Python Plotting Package}.
\newblock {\em Astronomical Data Analysis Software and Systems XIV ASP
  Conference Series,}, 347:91--95, 2005.

\bibitem[BQF{\etalchar{+}}01]{Barbieri2001}
R.~Barbieri, M.C. Quirk, L.M. Frank, M.A. Wilson, and E.N. Brown.
\newblock {Construction and analysis of non-Poisson stimulus-response models of
  neural spiking activity}.
\newblock {\em Journal of Neuroscience Methods}, 105(1):25--37, 2001.

\bibitem[BS06]{Bruno2006}
R.M. Bruno and B.~Sakmann.
\newblock {Cortex Is Driven by Weak but Synchronously Active Thalamocortical
  Synapses}.
\newblock {\em Science}, 312(5780):1622--1627, 2006.

\bibitem[BSK{\etalchar{+}}00]{Brenner2000}
N.~Brenner, S.P. Strong, R.~Koberle, W.~Bialek, and R.R.R. Steveninck.
\newblock {Synergy in a Neural Code}.
\newblock {\em Neural Computation}, 12(7):1531--52, 2000.

\bibitem[BT99]{Borst1999}
A.~Borst and F.E. Theunissen.
\newblock {Information theory and neural coding}.
\newblock {\em Nature Neuroscience}, 2:947--57, 1999.

\bibitem[CBNS98]{Castelo-Branco1998}
M.~Castelo-Branco, S.~Neuenschwander, and W.~Singer.
\newblock {Synchronization of Visual Responses between the Cortex, Lateral
  Geniculate Nucleus, and Retina in the Anesthetized Cat}.
\newblock {\em Journal of Neuroscience}, 18(16):6395--410, 1998.

\bibitem[EP75]{Eckhorn1975}
R.~Eckhorn and B.~Popel.
\newblock {Rigorous and extended application of information theory to the
  afferent visual system of the cat. II. Experimental results.}
\newblock {\em Biol Cybern}, 17(1):71--7, 1975.

\bibitem[FHM{\etalchar{+}}01]{Fellous2001}
J.M. Fellous, AR~Houweling, RH~Modi, RPN Rao, PHE Tiesinga, and TJ~Sejnowski.
\newblock {Frequency Dependence of Spike Timing Reliability in Cortical
  Pyramidal Cells and Interneurons}.
\newblock {\em Journal of Neurophysiology}, 85(4):1782--1787, 2001.

\bibitem[FNS07]{Fries2007}
P.~Fries, D.~Nikoli{\'c}, and W.~Singer.
\newblock {The gamma cycle}.
\newblock {\em Trends in Neurosciences}, 30(7):309--16, 2007.

\bibitem[Fre72]{Freeman1972}
WJ~Freeman.
\newblock {Measurement of oscillatory responses to electrical stimulation in
  olfactory bulb of cat}.
\newblock {\em Journal of Neurophysiology}, 35(6):762--779, 1972.

\bibitem[GKES89]{Gray1989}
C.M. Gray, P.~Koenig, A.K. Engel, and W.~Singer.
\newblock {Oscillatory responses in cat visual cortex exhibit inter-columnar
  synchronization which reflects global stimulus properties}.
\newblock {\em Nature}, 338(6213):334--337, 1989.

\bibitem[GT90]{Gelperin1990}
A.~Gelperin and D.W. Tank.
\newblock {Odour-modulated collective network oscillations of olfactory
  interneurons in a terrestrial mollusc}.
\newblock {\em Nature}, 345(6274):437--440, 1990.

\bibitem[HB66]{Heiss1966}
WD~Heiss and H.~Bornschein.
\newblock {Multimodal interval histograms of the continuous activity of retinal
  cat neurons}.
\newblock {\em Kybernetik}, 3(4):187--91, 1966.

\bibitem[HH52]{Hodgkin1952}
A.L. Hodgkin and AF~Huxley.
\newblock {Currents carried by sodium and potassium ions through the membrane
  of the giant axon of Loligo}.
\newblock {\em J. Physiol}, 116(4):449--472, 1952.

\bibitem[HY00]{Hutcheon2000}
B.~Hutcheon and Y.~Yarom.
\newblock {Resonance, oscillation and the intrinsic frequency preferences of
  neurons}.
\newblock {\em Trends in Neurosciences}, 23(5):216--222, 2000.

\bibitem[IGHT05]{Ishikane2005}
H.~Ishikane, M.~Gangi, S.~Honda, and M.~Tachibana.
\newblock {Synchronized retinal oscillations encode essential information for
  escape behavior in frogs}.
\newblock {\em Nat Neurosci}, 8(8):1087--95, 2005.

\bibitem[JM01]{Jarvis2001}
MR~Jarvis and PP~Mitra.
\newblock {Sampling Properties of the Spectrum and Coherency of Sequences of
  Action Potentials}.
\newblock {\em Neural Computation}, 13(4):717--49, 2001.

\bibitem[KFB57]{Kuffler1957}
SW~Kuffler, R.~Fitzhugh, and HB~Barlow.
\newblock {Maintained activity in the cat's retina in light and darkness.}
\newblock {\em J Gen Physiol}, 40(5):683--702, 1957.

\bibitem[KS05]{Koyama2005}
S.~Koyama and S.~Shinomoto.
\newblock {Empirical Bayes interpretations of random point events}.
\newblock {\em Journal of Physics A: Mathematical and General},
  38(29):L531--L537, 2005.

\bibitem[KTG{\etalchar{+}}04]{Kenyon2004}
G.T. Kenyon, J.~Theiler, J.S. George, B.J. Travis, and D.W. Marshak.
\newblock {Correlated firing improves stimulus discrimination in a retinal
  model}.
\newblock {\em Neural Computation}, 16(11):2261--91, 2004.

\bibitem[KWV{\etalchar{+}}08]{Koepsell2008}
K.~Koepsell, X.~Wang, V.~Vaingankar, Y.~Wei, Q.~Wang, D.L. Rathbun, W.M. Usrey,
  J.A. Hirsch, and F.T. Sommer.
\newblock {Retinal oscillations carry visual information to cortex}.
\newblock {\em submitted}, 2008.

\bibitem[LD94]{Laurent1994}
G.~Laurent and H.~Davidowitz.
\newblock {Encoding of Olfactory Information with Oscillating Neural
  Assemblies}.
\newblock {\em Science}, 265(5180):1872--75, 1994.

\bibitem[LV67]{Laufer1967}
M.~Laufer and M.~Verzeano.
\newblock {Periodic activity in the visual system of the cat.}
\newblock {\em Vision Res}, 7(3):215--29, 1967.

\bibitem[MC08]{Masse2008}
N.Y. Masse and E.P. Cook.
\newblock {The Effect of Middle Temporal Spike Phase on Sensory Encoding and
  Correlates with Behavior during a Motion-Detection Task}.
\newblock {\em Journal of Neuroscience}, 28(6):1343, 2008.

\bibitem[MHKS84]{Munemori1984}
J.~Munemori, K.~Hara, M.~Kimura, and R.~Sato.
\newblock {Statistical features of impulse trains in cat's lateral geniculate
  neurons}.
\newblock {\em Biological Cybernetics}, 50(3):167--172, 1984.

\bibitem[MJM{\etalchar{+}}08]{Muresan2008}
R.C. Muresan, O.F. Jurjut, V.V. Moca, W.~Singer, and D.~Nikolic.
\newblock {The Oscillation Score: An Efficient Method for Estimating
  Oscillation Strength in Neuronal Activity}.
\newblock {\em Journal of Neurophysiology}, 99(3):1333--53, 2008.

\bibitem[MRM{\etalchar{+}}08]{Montemurro2008}
M.A. Montemurro, M.J. Rasch, Y.~Murayama, N.K. Logothetis, and S.~Panzeri.
\newblock {Phase-of-Firing Coding of Natural Visual Stimuli in Primary Visual
  Cortex}.
\newblock {\em Current Biology}, 8(5):375--80, 2008.

\bibitem[Now97]{Nowak1997}
LG~Nowak.
\newblock {Influence of low and high frequency inputs on spike timing in visual
  cortical neurons}.
\newblock {\em Cerebral Cortex}, 7(6):487--501, 1997.

\bibitem[NS96]{Neuenschwander1996}
S.~Neuenschwander and W.~Singer.
\newblock {Long-range synchronization of oscillatory light responses in the cat
  retina and lateral geniculate nucleus}.
\newblock {\em Nature}, 379(6567):728--33, 1996.

\bibitem[OBL66]{Ogawa1966}
T.~Ogawa, P.~O. Bishop, and W.~R. Levick.
\newblock {Cortex Is Driven by Weak but Synchronously Active Thalamocortical
  Synapses}.
\newblock {\em Journal of Neurophysiology}, 29:1--30, 1966.

\bibitem[Oli07]{Oliphant2007}
T.E. Oliphant.
\newblock {Python for Scientific Computing}.
\newblock {\em Computing in Science \& Engineering}, 9(3):10--20, 2007.

\bibitem[OR93]{OKeefe1993}
J.~OÕKeefe and M.L. Recce.
\newblock {Phase relationship between hippocampal place units and the EEG theta
  rhythm}.
\newblock {\em Hippocampus}, 3(3):317--30, 1993.

\bibitem[PG07]{Perez2007}
F.~P{\'e}rez and B.E. Granger.
\newblock {IPython: A System for Interactive Scientific Computing}.
\newblock {\em Computing in Science \& Engineering}, 9(3):21--29, 2007.

\bibitem[PGM67]{Perkel1967}
D.H. Perkel, G.L. Gerstein, and G.P. Moore.
\newblock {Neuronal Spike Trains and Stochastic Point Processes: I. The Single
  Spike Train}.
\newblock {\em Biophysical Journal}, 7(4):391--418, 1967.

\bibitem[Rod67]{Rodieck1967}
R.W. Rodieck.
\newblock {Maintained activity of cat retinal ganglion cells}.
\newblock {\em Journal of Neurophysiology}, 5:1043--1071, 1967.

\bibitem[ROS90]{Richmond1990}
BJ~Richmond, LM~Optican, and H.~Spitzer.
\newblock {Temporal encoding of two-dimensional patterns by single units in
  primate primary visual cortex. I. Stimulus-response relations}.
\newblock {\em Journal of Neurophysiology}, 64(2):351--69, 1990.

\bibitem[RWvSB99]{Rieke1999}
F.~Rieke, D.~Warland, R.R. van Steveninck, and W.~Bialek.
\newblock {\em {Spikes: exploring the neural code}}.
\newblock MIT Press, Cambridge, 1999.

\bibitem[SBA03]{Szwed2003}
M.~Szwed, K.~Bagdasarian, and E.~Ahissar.
\newblock {Encoding of Vibrissal Active Touch}.
\newblock {\em Neuron}, 40(3):621--630, 2003.

\bibitem[SJL03]{Stopfer2003}
M.~Stopfer, V.~Jayaraman, and G.~Laurent.
\newblock {Intensity versus Identity Coding in an Olfactory System}.
\newblock {\em Neuron}, 39(6):991--1004, 2003.

\bibitem[SL03]{Sahani2003}
M.~Sahani and J.F. Linden.
\newblock {Evidence Optimization Techniques for Estimating Stimulus-Response
  Functions}.
\newblock In K.~Obermayer S.~Becker, S.~Thrun, editor, {\em Advances in Neural
  Information Processing Systems 15: Proceedings of the 2002 Conference}, pages
  109--16. MIT Press, 2003.

\bibitem[SZBB06]{Samonds2006}
J.M. Samonds, Z.~Zhou, M.R. Bernard, and AB~Bonds.
\newblock {Synchronous Activity in Cat Visual Cortex Encodes Collinear and
  Cocircular Contours}.
\newblock {\em Journal of Neurophysiology}, 95(4):2602--2616, 2006.

\bibitem[TFS08]{Tiesinga2008}
P.~Tiesinga, J.M. Fellous, and T.J. Sejnowski.
\newblock {Regulation of spike timing in visual cortical circuits}.
\newblock {\em Nature Reviews Neuroscience}, 9(2):97--107, 2008.

\bibitem[Tuc88]{Tuckwell1988}
H.C. Tuckwell.
\newblock {\em {Introduction to Theoretical Neurobiology}}.
\newblock Cambridge University Press, 1988.

\bibitem[UAR00]{Usrey2000}
W.M. Usrey, J.M. Alonso, and R.C. Reid.
\newblock {Synaptic Interactions between Thalamic Inputs to Simple Cells in Cat
  Visual Cortex}.
\newblock {\em Journal of Neuroscience}, 20(14):5461, 2000.

\bibitem[WWV{\etalchar{+}}07]{Wang2007}
X.~Wang, Y.~Wei, V.~Vaingankar, Q.~Wang, K.~Koepsell, F.T. Sommer, and J.A.
  Hirsch.
\newblock {Feedforward Excitation and Inhibition Evoke Dual Modes of Firing in
  the Cat's Visual Thalamus during Naturalistic Viewing}.
\newblock {\em Neuron}, 55(3):465--78, 2007.

\end{thebibliography}
\newcommand{\etalchar}[1]{$^{#1}$}

\end{document}